\documentclass[useAMS,usenatbib]{mn2e} % for a referee version

\usepackage{graphicx}
\usepackage{color}
\usepackage{mathtools}
\usepackage{txfonts}
\usepackage{epstopdf}
\newcommand{\gaia}{{\it Gaia }}

\title{Gaia Parallax of Milky Way Globular Clusters - A Solution of Mixture Model}

\author[Z. Shao and et. al. ]{
Zhengyi Shao$^{1,2}$ \thanks{E-mail: zyshao@shao.ac.cn (ZS)}
Lu Li$^{1,3}$
\\
        $^{1}$Shanghai Astronomical Observatory, Chinese Academy of Sciences, 80 Nandan Road, Shanghai 200030, People's Republic of China\\
        $^{2}$Key Lab for Astrophysics, 100 Guilin Road, Shanghai, 200234, People's Republic of China\\
        $^{3}$School of Astronomy and Space Science, University of Chinese Academy of Sciences, No.19A Yuquan Road, Beijing 100049, People's Republic of China\\
}
\date{Accepted 2019 August 13.
      Received 2019 August 1;
      in original form 2019 February 25}

\begin{document}

\date{\today}
\pagerange{\pageref{firstpage}--\pageref{lastpage}} \pubyear{2019}
\maketitle

\label{firstpage}
\begin{abstract}

We have established a mixture model approach to derive the parallax of the Milky Way globular clusters. It avoids the problem of cluster membership determination and provides a completely independent astrometrical solution by purely using the parallax data. This method is validated with simulated clusters of \cite{2017MNRAS.467..412P}. We have resolved 120 real globular clusters by the mixture model using parallaxes of the second data release of \gaia. They construct the largest direct parallax sample up to now. In comparing with other direct parallax results based on cluster members, including 75 clusters of \cite{2018A&A...616A..12G}, our method presents its accuracy, especially for some particular clusters. A systematic offset of $-27.6\pm1.7$ $\mu$as, together with a scatter of $22.8\pm1.3$ $\mu$as is found in comparing with other indirect parallax measurements. They are consistent with the global value and the variation of the zero-point of current \gaia parallaxes. Distances of several specific nearby globular clusters are discussed while the closest ones can reach high precisions, even taking the systematic error into account.

\end{abstract}


\section{Introduction}

Distance is one of the most fundamental parameters of the Milky Way globular clusters (hereafter GCs). Besides as a critical parameter for the spatial distribution and motion of GCs, it has a substantive impact on the researches of stellar populations, since distances of fiducial GCs could be the main source of uncertainty when using them to anchor stellar models.

GCs are distant objects, from 2 to $\sim$100 kpc away from us, so they are very difficult to have direct parallax measurements. Thus, indirect methods are widely used, such as the distance indicators of RR Lyrae, e.g., \cite{2016ApJ...827....2V,2019ApJ...871...49H} and contact binary, e.g., \cite{2013AJ....145...43K}, the main sequence subdwarf fitting, e.g., \cite{2017ApJ...838..162O}, the dynamical modeling, e.g., \cite{2015ApJ...812..149W,2018MNRAS.478.1520B} and the white-dwarf (WD) fitting, e.g., \cite{2012AJ....143...50W}. Besides the statistical uncertainty, they may also have systematic biases due to the stellar models, and/or be seriously influenced by the foreground dust reddening, which is difficult to be well detected.

The situation is now on the verge of dramatic improvement. Based on the high precision astrometrical observation of space facilities, the ability of direct parallax of GC is discussed with simulated data in \cite{2017MNRAS.467..412P} (hereafter P17). Measurements of nearby GCs are put into practice, such as \cite{2017AAS...22934307R} for M\,4, and \cite{2017ApJ...839...89W} for M\,4, NGC\,6397, M\,22, 47\,Tuc and M\,3 with the Tycho-Gaia Astrometric Solution catalog (TGAS), and \cite{2018ApJ...856L...6B} for NGC\,6397 with the data of Hubble Space Telescope (HST). Recently, with the second data release (DR2) of the \gaia mission \citep{2016A&A...595A...1G, 2018A&A...616A...1G}, more distant GCs reach into the scope of direct parallax. \cite{2018A&A...616A..12G} (hereafter H18) obtained the parallaxes ($\varpi_{\rm c}$) of 75 GCs ($\sim$half of the known Milky Way GCs) and used them as a tracer to discuss the zero-point and the calibration noise of \gaia parallax. \cite{2018ApJ...867..132C} determined the absolute \gaia parallax of two specific GCs (47\,Tuc and NGC\,362) based on their relative parallaxes to the Small Magellanic Cloud (SMC).

These works, including the simulation of P17, are all based on the cluster membership determination, either obtained through the proper motions or further checked with the colour-magnitude diagram (CMD). There is a dilemma of such {\it membership} approaches. A {\it strict} criterium may rule out most of the cluster members. It will lose the statistical benefits of the sample size, e.g. the limited members with large errors ($\sigma_{\varpi}$) may lead to a remarkable uncertainty of the cluster parallax ($\sigma_{\varpi_{\rm c}}$). On the other hand, a {\it loose} criterium cannot avoid the contamination of field stars. This influence could be serious because $\sigma_\varpi$ of individual stars significantly depend on their distances and brightness, while a few contaminated foreground and/or bright background stars with smaller errors may strongly bias the mean parallax of the cluster.

Alternatively, if we only focus on the mean parallax of a cluster, it is possible to employ the mixture model to fit the cluster parallax as a parameter directly, while tolerating the existence of an additional component of the field stars. It can avoid the problem of membership determination, and take the utilisation of the survey data as much as possible. In this work, we try to establish a mixture model solution for the parallax of GC that only using the parallax data themselves and to examine how many GCs can be resolved with the \gaia DR2.

Same as many other space missions, it is found that there are zero-points of the \gaia parallaxes, with the global average value $\simeq-0.029$ mas and the variation $\sim0.025$ mas \citep{2018A&A...616A..17A, 2018A&A...616A..12G}, which is mainly due to the calibration bias and noise. It will dominate the systematic error of distant GC. However, a pure direct parallax measurement is still important, since it is independent of any other observational data and stellar models and will provide an essential reference for all other distance measurements. Moreover, the mean parallaxes of a distant GC may have higher statistical precision than that of individual stars, so it is worthwhile in estimating the zero-point and its variations of the \gaia parallaxes if as many GCs as possible could be involved. Therefore, we expect that the mixture model solution can provide a unique approach, which will diversify the era of the direct parallax of GC.

The algorithm of the mixture model and the validation with simulated clusters are described in section \ref{sec:algorithm}. The parallax data and fitting results for real GCs are explained in section \ref{sec:result}. Comparisons and discussions are presented in section \ref{sec:discussion}, and conclusions are briefly summarized in section \ref{sec:summary}.

\begin{figure}
\vspace*{-1.0cm}\hspace*{-0.50cm}\includegraphics[width=1.2\columnwidth]{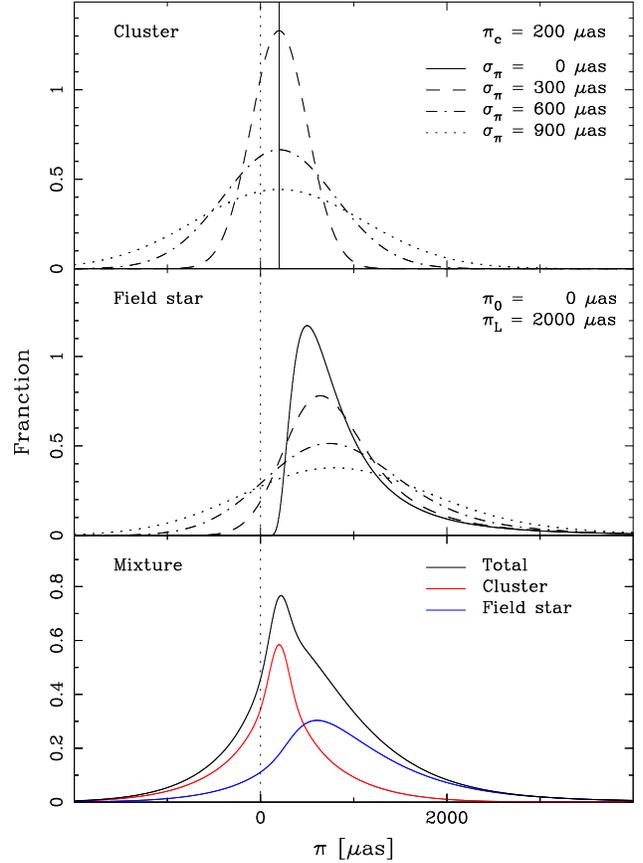}
\vspace*{-0.5cm} \caption{Model distribution of parallaxes. The top and middle panels are for the cluster members and the field stars separately. The solid lines show the intrinsic distributions of $\phi_{\rm c}$ and $\phi_{\rm f}$. The dashed, dash-dotted and dotted lines show the apparent distributions of $\varphi_{\rm c}$ and $\varphi_{\rm f}$ with observational errors of $\sigma_{\varpi} = 300,600,900$ $\mu$as respectively. The bottom panel shows the combinations of the cluster members (red line) or the field stars (blue line) with their $\sigma_{\varpi}$ of individual stars uniformly distributed from 100 $\mu$as to 1.0 mas. The black line shows the mixture of these two components with equal star numbers.    } \label{Fig1}
\end{figure}

\begin{figure*}
\vspace*{-2.0cm}\hspace*{-3.50cm} \includegraphics[scale=.90]{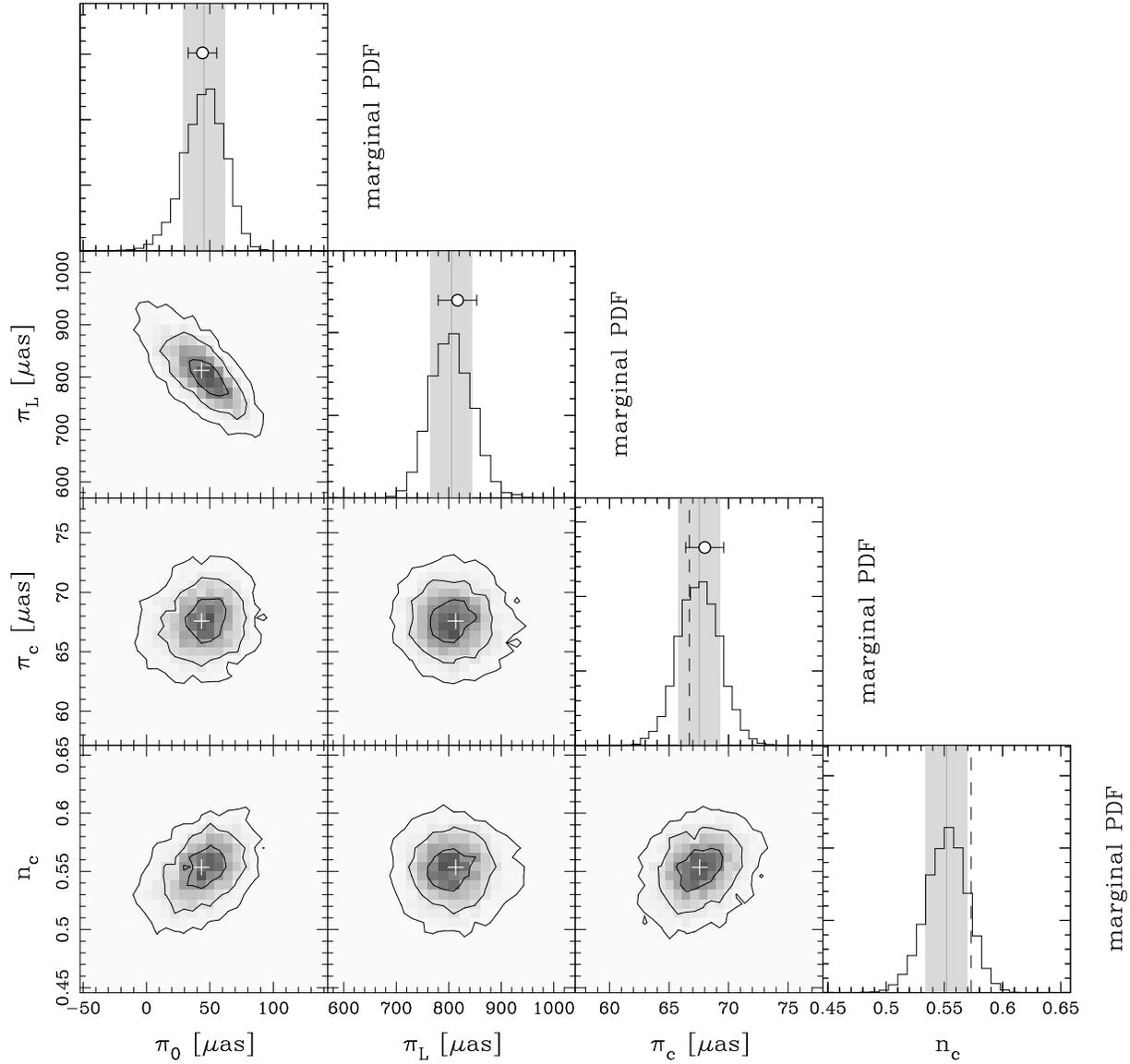}
\vspace*{-1.5cm} \caption{Probability density functions (PDFs) of parameters based on the Nested Sampling approach in the fitting of a simulated GC at $d=15$kpc with contamination of {\it disc} field stars (GC\,06 of P17). Sample stars are selected within 3$r_h$. White cross symbols indicate the maximum likelihood point. Histograms show the 1-dimensional marginalized PDFs for each parameter, with vertical grey lines and shadows represent the mean values and the standard deviations. Hollow circles with horizonal error bars indicate the results of the independent fittings for pure cluster members ($\varpi_{\rm c}$) or field stars ($\varpi_0, \varpi_L$) respectively. The true values of $\varpi_{\rm c}$ and $n_{\rm c}$ are shown as dashed vertical lines.} \label{Fig2}
\end{figure*}

\section{Model and Algorithm}\label{sec:algorithm}

The mixture model is a natural statistical method for many situations in astronomy \citep{2017arXiv171111101K}. It combines multiple components into a single density profile in a multi-dimensional phase space of the observational data. In this work, we dedicate in the data of parallax and build the mixture model for only two components, the cluster members and the field stars within the projected area centred on a GC.

\cite{2018A&A...612A..99R} employed a mixture model of two Gaussians for the \gaia DR1 parallax of open clusters (see figure 5 of their paper). This simple assumption was feasible for nearby clusters since those field stars are almost backgrounds leading to an obvious bimodal feature. But, in the cases of distant GC, the intrinsic distributions of neither the field stars nor the cluster members are Gaussian. These two components will be heavily mixed in the place of small parallax, and the larger observational errors enhance the mixture degree. Therefore, a rigorous model or solution should be considered.

On the other hand, the GCs have their favorableness. Comparing to the distance, the size of a cluster is neglectable. So its intrinsic distribution could be treated as a $\delta$ function. Moreover, the large number of sample stars will ensure the precision of the mean value of the cluster, though the individual stars have large errors.

\subsection{Mixture model for the parallax distribution}\label{sec:mm}

The cluster members and the field stars are assumed to have a mixture distribution of parallax ($\rm{\varpi}$) intrinsically:
%--
\begin{align}\label{eq-varPhi}
% \nonumber % Remove numbering (before each equation)
    \varPhi(\varpi) &= \varPhi_{\rm c}(\varpi) + \varPhi_{\rm f}(\varpi)  \nonumber               \\
                    &= n_{\rm c} \phi_{\rm c}(\varpi) + n_{\rm f}\phi_{\rm f}(\varpi)
\end{align}
%--
\noindent where $\phi_{\rm c}$ and $\phi_{\rm f}$ are normalized distributions of the cluster members and the field stars respectively. $n_{\rm c}$ and $n_{\rm f}$ are their fractions with $n_{\rm c} + n_{\rm f} =1 $. The $\phi_{\rm c}$ is assumed as a $\delta$ function,
\begin{equation}\label{eq-GC}
    \phi_{\rm c}(\varpi)=\delta(\varpi - \varpi_{\rm c}),
\end{equation}
\noindent with $\varpi_{\rm c}$ to be the parallax of the cluster. The $\phi_{\rm f}$ could be approximated as a function derived from the simple exponentially decreased number density profile of eq.(1) of \cite{2018AJ....156...58B},
%--
\begin{equation}\label{eq-FS}
\phi_{\rm f}(\varpi)=\begin{dcases}
    \frac{\varpi_L^3}{2(\varpi-\varpi_0)^4}\exp {\left(\frac{-\varpi_{L}}{\varpi-\varpi_0}\right)}, & {\rm if}~~\varpi>\varpi_0, \\
   0                          & {\rm otherwise}
\end{dcases}
\end{equation}
%--
\noindent where $\varpi_L$ is the reciprocal of the scale length $L$ and determines the width and also the peak (at $\varpi_L/4$) of the profile. $\varpi_0$ is introduced as the zero-point that indicates the lower limit of parallax in a given direction. Since eq.~\ref{eq-FS} is an approximation, so we caution here that the $\varpi_0$ can only be regard as a nominal zero-point, which cannot be used to analyze the \gaia zero-point directly (see Sec.~\ref{sec:validation} and figure ~\ref{Fig2} for details).

Then, we have totally four model parameters, $n_{\rm c}$, $\varpi_{\rm c}$, $\varpi_L$ and $\varpi_0$. They curve the intrinsic distribution of the mixture of these two components.
%--
\begin{figure}
\vspace*{-1.0cm}\hspace*{-0.50cm}\includegraphics[width=1.2\columnwidth]{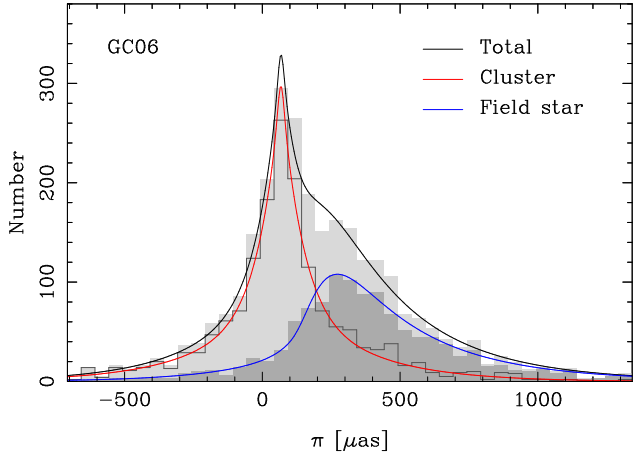}
\vspace*{-0.5cm}\caption{Histograms of parallax distributions of the simulated cluster of GC\,06 of P17. The cluster members, the field stars and their mixture are shown as dark grey line, grey shadow and light grey shadow separately. Corresponding apparent model distributions of these two components and their mixture are also shown for comparison. } \label{Fig3}
\end{figure}
%--

For the current \gaia parallax, even it has reached the highest precision up to now, the errors of most sample stars are still very large, which cause the apparent distributions of both of the cluster and the field widely extended and sufficiently mixed.

For individual stars with a given observational error $\sigma_{\varpi}$, they may follow the apparent distribution $\psi$ that convolving $ \phi$ with a Gaussian kernel having the standard deviation of $\sigma_{\varpi}$. Then we have
%--
\begin{equation}\label{eq-psiX}
    \psi_{\rm x}(\varpi,\sigma_{\varpi}) = \int_{}^{} \phi_{\rm x}(\varpi')\cdot \mathcal{N}(\varpi';\varpi,\sigma_{\varpi}) \,d\varpi' ,
\end{equation}
%--
\noindent where $\mathcal{N}(\varpi';\varpi,\sigma_{\varpi})$ represents the Gaussian probability centred at $\varpi$ and the subscript 'x' infers 'c' or 'f' for the cluster or the field respectively. Practically, if we consider a truncated data set, e.g., constrain the parallax range of $[\varpi_{min},\varpi_{max}]$ to exclude some outliers, we should also re-normalize $\psi_{\rm x}$ by the factors of
%--
\begin{equation}\label{eq-C}
    \mathcal{C}_{\rm x}(\sigma_{\varpi}) = \left[ \int_{\varpi_{min}}^{\varpi_{max}} \psi_{\rm x}(\varpi,\sigma_{\varpi}) \,d\varpi \right]^{-1}.
\end{equation}
%--
\noindent Thus, for the $i$th star with observational parallax ($\varpi_i$) and error ($\sigma_{\varpi_i}$), it obeys the mixture apparent distribution of
%--
\begin{align}\label{eq-varPsi}
    \varPsi(\varpi_i,\sigma_{\varpi_i}) &= \varPsi_{\rm c}(\varpi_i,\sigma_{\varpi_i}) + \varPsi_{\rm f}(\varpi_i,\sigma_{\varpi_i})   \nonumber       \\
               &= n_{\rm c}\, \mathcal{C}_{\rm c}(\sigma_{\varpi_i}) \,\psi_{\rm c}(\varpi_i,\sigma_{\varpi_i}) + n_{\rm f} \, \mathcal{C}_{\rm f}(\sigma_{\varpi_i}) \,\psi_{\rm f}(\varpi_i,\sigma_{\varpi_i}) .
\end{align}
%--

Figure~\ref{Fig1} shows the model distributions of parallax, both for the intrinsic and the apparent ones, of cluster members, field stars and their mixture in separate panels. Clearly, for the cluster, all profiles are symmetrical with their wings extended based on the errors. For the field, it is significantly asymmetrical with a longer wing at the right side (at larger $\varpi$). With the increase of the error, the shape also extends and tends to have less asymmetry. Meanwhile, the peak positions slightly shift towards right. Regardless, the asymmetry always exists, and this is the main feature that distinguish the field stars to the cluster. Therefore, the purpose of our mixture model is to derive the $\varpi_{\rm c}$ as a fitting parameter from such heavily contaminated apparent distribution. On the other words, we are trying to peel out the symmetric cluster distribution from the mixed asymmetric one.

It should be mentioned that, when we fit such kind of error-dominated distribution, the accuracy of results depends on the credibility of the observational errors. For each source published in \gaia DR2, it is derived from a simultaneous five-parameter fit of an astrometric source model to the data, and thus comprise five astrometric parameters with their associated uncertainties, also with ten correlation coefficients between the estimated parameters. In this work, since we only use the parallax, so we do not need to take the covariances into account. What we should concern is if the \gaia DR2 overestimates or underestimates the parallax errors. According to the analysis of \cite{2018A&A...616A...2L}, a factor of 1.081 may be employed to all $\sigma_{\varpi}$, though we found it is too small to affect the conclusions of this paper.

In fitting of the model parameters, we assume that the likelihood of the $i$th star following the apparent distribution with $\sigma_{\varpi_i}$,
%--
\begin{equation}\label{eq-lh}
    \mathcal{L}_i = \varPsi(\varpi_i,\sigma_{\varpi_i}) ,
\end{equation}
%--
\noindent then write down the joint logistical likelihood for the whole sample as
%--
\begin{equation}\label{eq-lhtot}
    \ln \mathcal{L} = \sum \ln \mathcal{L}_i .
\end{equation}
%--

We employ the Nested Sampling method \citep{2013arXiv1306.2144F} to map the full probability density function (PDF) of four model parameters, $n_{\rm c}$, $\varpi_{\rm c}$, $\varpi_L$ and $\varpi_0$. Since we only concern the value of $\varpi_{\rm c}$, we subsequently  marginalize over the other three parameters. The marginal PDF is theoretically integrated from all probabilities of other uninteresting parameters, and under the Bayesian framework, it is a rigorous way if we are interesting in only one parameter. Also, it is expected to be generally broader than the conditional PDF at the best fitting point. Then, we use the marginal PDF to derive its mean value and standard deviation as the final fitting result and error of $\varpi_{\rm c}$ (see figure~\ref{Fig2} as an example).
%--
\begin{figure*}
\vspace*{-2.5cm}\hspace*{-0.5cm} \includegraphics[scale=.70]{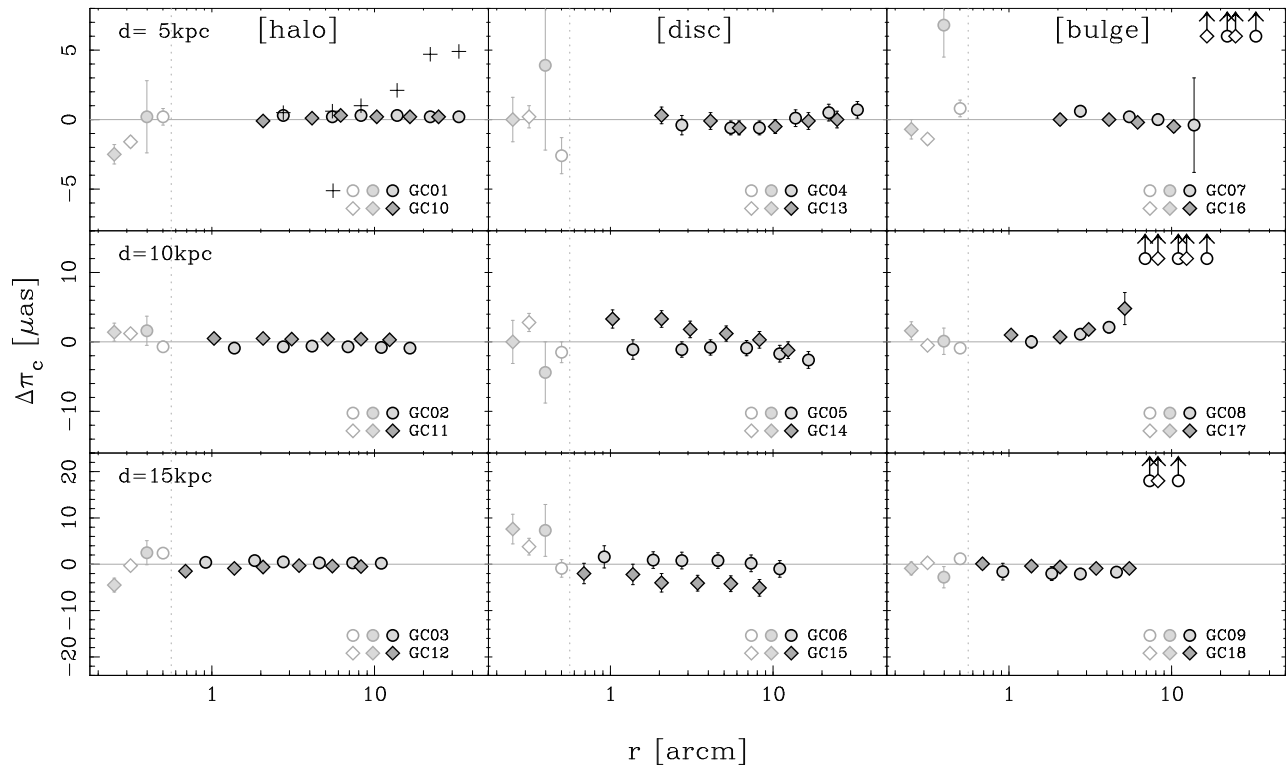}
\vspace*{-2.0cm} \caption{Comparison of 18 simulated GCs of P17. $\Delta\varpi_{\rm c} = \varpi_{\rm c,fitting} - \varpi_{\rm c,true}$ is the difference between the fitting result and the true input value.  GCs at 5, 10 and 15 kpc are lined from top to bottom, and different contaminations of the {\it halo}, {\it disc} and {\it bulge}, are arranged from left to right. Different concentrations, $c=1$ (GC 1 to 9) or 2.5 (GC 10 to 18), are plotted in the same sub-panel with circles and quadrangles respectively. The selected cluster ranges are 1, 2, 3, 5, 8 and 12 times of $r_h$. In cases of the {\it bulge} GCs, some subsamples of larger radii that cannot be resolved with the mixture model are plotted as hollow symbols with up arrows at the up-right corners of each sub-panel. As a comparison, for GC01, the mean parallaxes fitted by a single Gaussian model are plotted as plus symbols. The results of the {\it membership} approach (P17) are plotted in grey colour on the left side of the dotted lines of each sub-panel, with hollow symbols for the {\it loose} member and filled symbols for the {\it strict} member cases.} \label{Fig4}
\end{figure*}
%--

\subsection{Validation with simulated clusters}\label{sec:validation}

P17 generated 18 simulated GCs with two concentrations of $c=log(r_t/r_c)=1 ~{\rm or} ~2.5$, three distances at 5, 10 or 15 kpc and three background contaminations of the {\it halo}, {\it disc} or {\it bulge} conditions. They computed the cluster parallaxes with selected GC members for a {\it loose} criterium of 1 mas yr$^{-1}$ or a {\it strict} criterium of 0.3 mas yr$^{-1}$ from the cluster proper motion. They found that the differences between the true input parallaxes and the recovered ones are  about $1\%$ to $5\%$, and the formal errors are of the same order. They also claimed that choosing a more restrictive membership selection can often increase the formal errors without resulting in a better $\varpi_{\rm c}$ determination, and on the contrary, the bias is slightly increased.

Here we use the mixture model method to derived $\varpi_{\rm c}$ for these simulated GCs. For example, figure~\ref{Fig2} is the Nested Sampling calculated PDF of parameters for a simulated cluster, GC\,06. One can see that, all four parameters converge very well with pretty small uncertainties and have fairly symmetric shapes of PDF.

As a comparison, we also fit these two components independently based on the sub-samples of the cluster or the field, with the results shown as hollow circles with horizontal error bars. It is found that the results of $\varpi_0$, $\varpi_L$ (for the field) and $\varpi_{\rm c}$ (for the cluster) are well consistent with those corresponding values from the mixture model, which can verify the validity and rationality of our approach. Also, $\varpi_{\rm c}$ is in good agreement with the true input value (the dashed line) of this GC, while $n_{\rm c}$ is slightly underestimated but still in agreement when considering the uncertainty.

We also notice that there is a significant correlation between $\varpi_0$ and $\varpi_L$. Moreover, one may find that the value of $\varpi_0$ does not verge on zero, though there is no zero-point assumed for these simulated stars and even if we fit the field component independently. That is because the real (or simulated) distribution of the field stars is also complicated, i.e., a more detailed distribution should include at least components of the disc, the bulge and the halo. When we use the simple approximation of eq.~\ref{eq-FS} in the fitting, the value of $\varpi_0$ will sensitively offset according to the difference from the real distribution. For this reason, we can not use the fitting result of $\varpi_0$ to discuss the \gaia zero-point directly, and a detailed modelling of the field stars is beyond the scope of this paper. Fortunately, $\varpi_{\rm c}$ is free from this effect and appears to have a rigorous and accurate value.

Figure~\ref{Fig3} is a comparison of the parallax distribution for this GC with its best-fitting model. In order to produce the apparent model distributions, the intrinsic model should be convolved with the observational errors that following the allocation of errors of those simulated stars. For instance, suppose there are sufficient number of model stars, $N_s$, with $n_{\rm c}N_s$ cluster members and $n_{\rm f}N_s$ field stars, we then randomly pick up errors from the whole simulated sample of GC\,06. Then we calculate $\varPsi_{\rm c}$ or $\varPsi_{\rm f}$ for each model star and finally sum them up as the apparent model distribution of the cluster (the red curve) or the field (the blue curve) separately. It is interesting that even if we have not used any prior membership information from the simulated data, the model curves match the histograms of both of them very well, either in the shapes or in the amplitudes. Certainly, the model curve of their mixture (the black curve) also match the whole sample (the light grey histogram) well and behaves the asymmetric feature properly.

For all 18 simulated GCs of P17, we apply this method for stars within different cluster radii covering a quite broad range, from 1 to 12 times of the $r_h$. Fitting results are plotted in figure~\ref{Fig1}, together with those of P17 for comparison.

Generally, the fitting process converges for almost all of the selected radii except for some larger radii of the {\it bulge} GCs, which suffer too much contamination of field stars to recognise the cluster. As shown in figure~\ref{Fig4}, all converged results are similar to or even better than the best cases of P17. Also, our results are stable in a wide range of the variation of cluster size, which can guarantee the flexibility in choosing the fitting size for real clusters. Therefore, it can be concluded that the mixture model is demonstrated to be a practical and robust method in determining the mean parallax of GC.

Moreover, as a comparison, we have tried the single Gaussian distribution, other than the mixture model, for the simulated GC01, which is the most {\it easy} case of the {\it halo} clusters. It is under the assumption that all sample stars are cluster members. The results can be found in the top-left panel of figure \ref{Fig1} as the plus symbols. The biases of $\varpi_{\rm c}$ systematically increased (to smaller distance) with enlargement of the cluster radius, within which more field stars are included, though they are at most $\sim 1\%$. In fact, these several field stars just {\it require} a more disperse distribution in order to eliminate their influence on the cluster. It reveals the key concept of the mixture model that we do not need to rule out the field stars, but give them an appropriate extended distribution, while the simple exponentially decreased profile of eq.\ref{eq-FS} seems acceptable when the field stars do not dominate the sample.

\section{Results of gaia parallax}\label{sec:result}

%--
\begin{table}
  \caption{Fitting results of parallax of globular clusters}\label{table1}
\begin{center}
\begin{tabular}{llcrrrr}
\hline \hline
  \multicolumn{1}{c}{Id} &
  \multicolumn{1}{c}{Name} &
%  \multicolumn{1}{c}{Range} &
  \multicolumn{2}{c}{Radius} &
  \multicolumn{1}{c}{$N_{\rm s}$} &
  \multicolumn{1}{c}{$\varpi_{\rm c}$} &
  \multicolumn{1}{c}{$\sigma_{\varpi_{\rm c}}$} \\
 & & & $arcm$ && $\mu as$ & $\mu as$\\
\hline
NGC\,6121 & M\,4 & $r_1$ & 28.2 & 59388 & 499.5 & 1.1\\
NGC\,6397 &  & $r_1$ & 15.3 & 46027 & 382.4 & 1.3\\
NGC\,5139 & $\omega$\,Cen & $r_1$ & 27.0 & 114921 & 136.8 & 1.5\\
NGC\,104 & 47\,Tuc & $(r_0+r_1)/2$ & 21.9 & 98410 & 195.1 & 0.5\\
NGC\,6266 & M\,62 & $r_1$ & 7.8 & 24354 & 145.1 & 3.1\\
\hline
\end{tabular}
\end{center}
\raggedright{Note: (1) $\varpi_{\rm c}$ is the apparent parallax of GC without the correction of zero-point. (2) $\sigma_{\varpi_{\rm c}}$ is only the statistical error while a significant systematic uncertainty should also be considered (see \ref{sec:discussion2} for details). (3) This table is available in its entirety in the electronic version. }
\end{table}
%--

For the real GCs, we take the \gaia DR2 parallaxes of stars within a given radius of the cluster. Practically, we adopt a flux limitation of $G=20.7$ mag, which is about the photometric completeness level of the current \gaia data, and modify the parallax error of stars $\sigma_{\varpi}$ with a factor of 1.081 \citep{2018A&A...616A...2L}. We further constrain stars within $-5 < \varpi <10 $ mas and exclude a few outliers with $ \varpi + 0.029 ~{\rm mas} < -3\sigma_{\varpi}$.  It should be mentioned that, the real GCs are crowding fields so in the very central region it may be blended and the sample is incomplete, like $\omega$\,Cen and 47\,Tuc. However, since we only discuss the parallax distribution, such kind of radial and magnitude dependent incompleteness will have no relevant effect on the $\varpi_{\rm c}$ result, but will modify the fraction of the cluster, $n_{\rm c}$.

We have tried all 147 GCs listed in \cite{2013A&A...558A..53K} and finally found 120 of them can be reasonably resolved by the mixture model. These clusters must satisfy two constraints. One is that in using the Nested Sampling, the marginal PDF of $\varpi_{\rm c}$ is a single-mode shape and the most probable value is similar to its average. Another constraint is that the $\varpi_{\rm c}$ results appear to be stable in a proper wide range of size that covers the given radius.

The given radius of a cluster is chosen orderly from three characteristic values of $r_1$, $(r_0+r_1)/2$ or $r_0$, with $r_0$ and $r_1$ to be the visible radii for the core and the central part of a cluster \citep{2012A&A...543A.156K}. Among them, $r_1$ is the first choice since more sample stars will result in higher precision, and 71 of 120 GCs are resolved with this radius. For other GCs, mostly more distant and contaminated, we choose one of the other smaller radii based on their over density areas and the stability of fitting results.

\begin{figure*} %[ht]
\vspace*{-3.0cm}\hspace*{-0.50cm} \includegraphics[width=2.3\columnwidth]{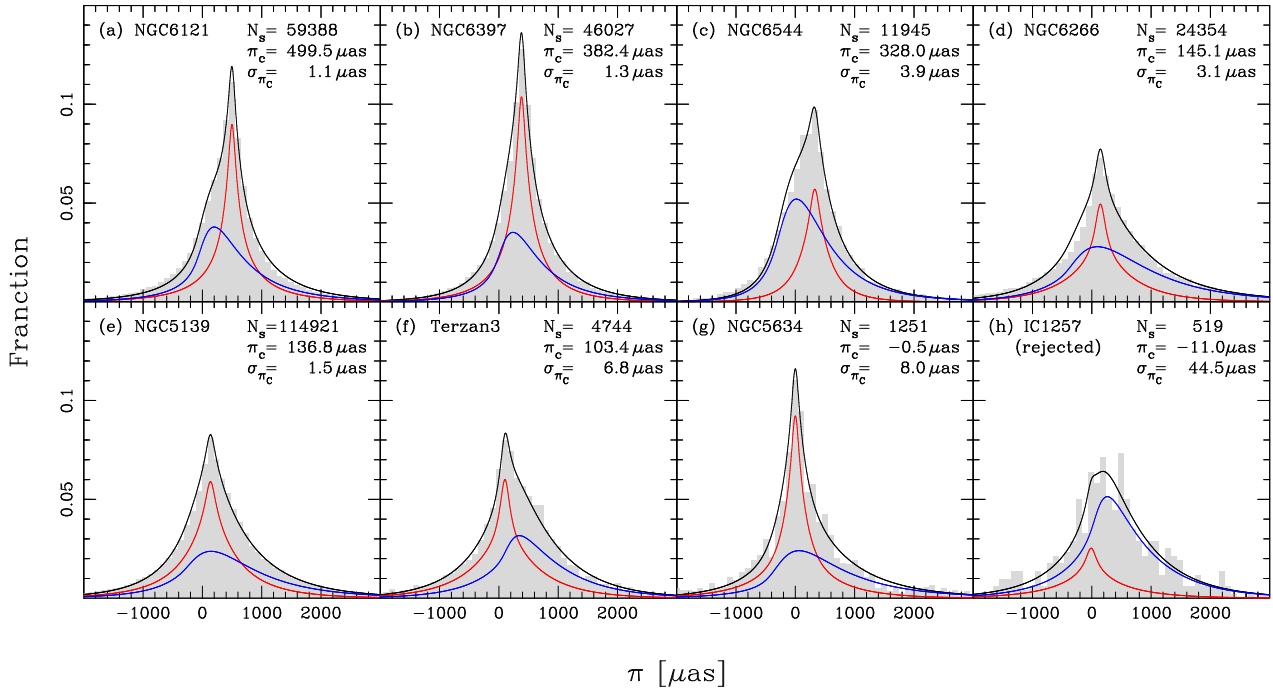}
\vspace*{-2.5cm}\caption{Histograms of parallax distributions of selected GCs. Model distributions of the cluster (red), the field (blue) and their mixture (black) are shown for comparison. Panels (a)-(g) are GCs at different distances, with which can show the various of the relationships of these two components. Panel (h) is an example of a rejected cluster with unacceptable fitting.} \label{Fig5}
\end{figure*}

Figure~\ref{Fig5} shows some examples of the fittings. It is the same as figure~\ref{Fig3}, but without the independent histograms of the cluster or the field. For seven examples of finally resolved clusters, their model curves (black curve) match the observational data very well, whatever $\varpi_{\rm c}$ is larger or smaller than the peak value of the field, or in some cases, the cluster does not overwhelmingly dominate the distribution. Panel (h) is one of the 27 rejected clusters, while in this case, it has not enough sample stars and the cluster component is too weak to give us believable fitting results.

Fitting results of $\varpi_{\rm c}$ of 120 GCs, together with their chosen radii are listed in table~\ref{table1}. Typically, $\varpi_{\rm c}$ have high fitting precision with the median value of $\sigma_{\varpi_{\rm c}} \simeq 6.8$ $\mu$as, while the most precise value is 0.5 $\mu$as. The median value of the fractional uncertainty $f = \sigma_{\varpi_{\rm c}}/\varpi_{\rm c} \sim 0.06 $ for the whole sample. There are 9 nearby GCs with $f \lesssim 0.01$, which means that they are better than the 1$\%$ precision level, if only in terms of the statistical error. Meanwhile, it is obvious that as the distance increases or the number of sample stars decreases, the precision decreases.

Overall, by using the mixture model method, we obtain the largest sample of GCs with direct parallax up to now. However, it is worth emphasizing the fact that the fitting result of $\varpi_{\rm c}$ is only the apparent mean value of the cluster, and should be corrected the zero-point to obtain its absolute parallax.
%--
%--

\section{Comparisons and Discussions}\label{sec:discussion}

\subsection{Compare with other direct parallaxes}\label{sec:discussion1}

\begin{figure}
\vspace*{-1.0cm}\hspace*{-0.50cm}\includegraphics[width=1.2\columnwidth]{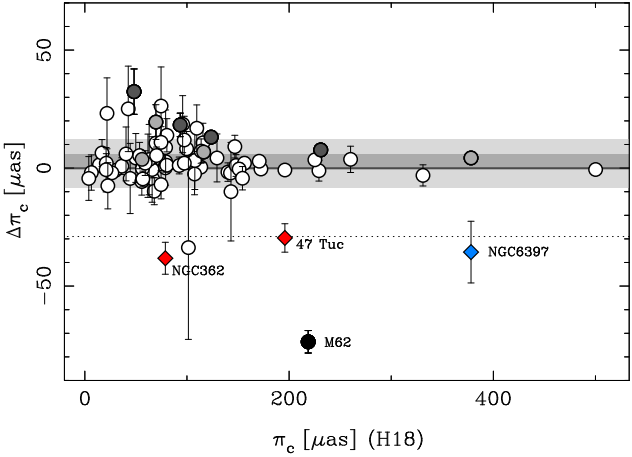}
\caption{Comparison of the mean parallax of GCs between this work (TW) and other direct parallax results. $\Delta\varpi_{\rm c} = \varpi_{\rm c,TW} - \varpi_{\rm c,others}$, and the error bars $\sigma_{\varpi_{\rm c,syn}}$ represent the quadratic sum of both of their uncertainties. Circles are 75 GCs from H18, with grey filled for $|\Delta\varpi_{\rm c}| > 2\sigma_{\varpi_{\rm c,syn}}$ and dark grey filled for $|\Delta\varpi_{\rm c}| > 3 \sigma_{\varpi_{\rm c,syn}}$. Red quadrangles are from Chen et al. (2018) and the blue quadrangle is from Brown et al. (2018). Light grey shadow shows the scatter (1$\sigma$ range) centered on the average value of $\Delta\varpi_{\rm c}$, while the narrower grey shadow has the same meaning but exclude M\,62(NGC\,6266)   (see table~\ref{table2} for these values). The dotted line presents the $-29$ $\mu$as offset of the global zero-point of \gaia DR2.} \label{Fig6}
\end{figure}

H18 presented parallaxes of 75 GCs based on their proper motion members. All of them are resolved by our mixture model. Generally, for these common GCs, values of $\varpi_{\rm c}$ are in good agreement, with the difference, $\Delta\varpi_{\rm c} = \varpi_{\rm c,TW} - \varpi_{\rm c,H18}$, to be only a few $\mu$as, which is neglectable especially for some nearby clusters. It could be found in table \ref{table2} and figure \ref{Fig6} for details, while in computing the $\langle \Delta\varpi_{\rm c} \rangle$ and the scatter $\sigma_{\Delta\varpi_{\rm c}}$, the fitting errors of individual GCs from these two methods are taken into account and synthesized by their quadratic sum $\sigma_{\varpi_{\rm c,syn}}^2 = \sigma_{\varpi_{\rm c,TW}}^2 + \sigma_{\varpi_{\rm c,H18}}^2 $.

On the other hand, if we compare $\Delta\varpi_{\rm c}$ to the  $\sigma_{\varpi_{\rm c,syn}}$, one may find that 9 or 5 of 75 GCs exceed the 2$\sigma$ or 3$\sigma$ levels. These ratios, $12\% ~{\rm or}~ 7\%$, are too large to satisfy the expectations of the Gaussian distribution, which are $\lesssim 5\%$ or $\lesssim 0.3\%$ respectively. Although the small sample size statistics could be considered as one reason, this discrepant also implies the method dependence and recall the importance of the diversity of direct parallax approach.

The most difference appears in M\,62 (NGC\,6266). The $\varpi_{\rm c,H18}=218.7\pm3.6$ $\mu$as, is much larger than the value of this work $\varpi_{\rm c,TW}=145.0\pm3.0$ $\mu$as. Comparing with \cite{1996AJ....112.1487H} (hereafter H10), $\varpi_{\rm c,H18}$ is also much larger than its expectation as shown in figure 6 of H18. We suspect it probably dues to the contamination of remaining field stars. Since this cluster is projected on an extremely dense field in the direction of the Galactic bulge, and unfortunately its proper motion is not clearly separated to those of field stars, so it is impossible to exclude most of the foreground/background stars. In contrast, $\varpi_{\rm c,TW}$ appears without any speciality in comparing with other indirect measurements (see the filled black symbols in figure \ref{Fig7}). Taking this cluster as an example, it presents the advantage of the mixture model of avoiding the membership determination.

On average, the fitting uncertainties of this work are a little bit larger than those of H18. It may have statistical origins. One is that even if there are similar numbers of sample star of a given cluster, $\sigma_{\varpi_{\rm c,TW}}$ is obtained from the marginal PDF that involves the correlations between $\varpi_{\rm c}$ and all other parameters, and will be generally broader. It will then have a larger value of uncertainty, which is reasonable in the view of the Bayesian framework. Moreover, H18 did not account the bias caused by the possible contamination of field stars, so their precisions are higher and simply increase with the number of sample star. In other words, it could be regarded that the mixture model is a tradeoff at the expense of a little precision but increases the accuracy.

In figure \ref{Fig6}, we also plot three individual clusters with other direct parallaxes. 47\,Tuc (NGC\,104) and NGC\,362 are measured by \cite{2018ApJ...867..132C}, also with the \gaia DR2. They used the pairwise method to obtain the relative parallaxes between some cluster members and SMC stars and then corrected them to the absolute parallaxes based on the distance of SMC. NGC\,6397 is measured by \cite{2018ApJ...856L...6B} with the parallax data of HST of 39 selected cluster members. For all of these three GCs, their parallaxes are consistent with our results at an offset of the global zero-point of $\sim 29$ $\mu$as of \gaia DR2, whereas their uncertainties are much larger since they only can use limited members, which result in less precision.

In summary, we can conclude that the mixture model method provides a practical approach in measuring the direct parallax of GC.
%--
%--

\subsection{Compare with indirect parallax methods}\label{sec:discussion2}

%--
\begin{table}
\begin{center}
  \caption{Comparison of parallaxes of globular clusters}\label{table2}
  \begin{tabular}{crrcc}
  \hline\hline
    Source &  $N_{GC}$ & $N_{\Delta\varpi_{\rm c}}$ & $\langle \Delta\varpi_{\rm c} \rangle$ & $\sigma_{\Delta\varpi_{\rm c}}$ \\
   \hline
    \multicolumn{5}{c}{\it compare with direct parallax} \\
   \hline
   H18 & 75 & 75 & $~~~~2.0\pm1.4$ & $10.3\pm1.2$\\
   (excl. M\,62 ) & 74 & 74 & ~~~~$2.4\pm0.7$ &  $~3.3\pm0.7$\\
  \hline
    \multicolumn{5}{c}{\it compare with indirect parallax}\\
   \hline
   H10  &120 &120 & $-27.2\pm2.2$ & $21.1\pm1.8$  \\
  post-H10  & 66 & 118 & $-30.4\pm2.4$ & $24.2\pm1.8$  \\
   Combined & 120 & 238 & $-27.6\pm1.7$ & $22.8\pm1.3$  \\
   \hline
  \end{tabular}
\end{center}
\raggedright{Note: $\Delta\varpi_{\rm c} = \varpi_{\rm c,TW} - \varpi_{\rm c,others}$.}
\end{table}
%--

We compare our results with indirect parallax measurements from multiple sources. One is from H10 who collected the data of all known GCs at that time. Their distance sources are mixed of multiple methods but mainly based on the indicator of the horizontal branch. For recent results (hereafter post-H10), we compare with three other primary methods having relatively bulk of data, the dynamical modelling (15 GCs from \cite{2015ApJ...812..149W} and 53 GCs from \cite{2018MNRAS.478.1520B}), the main-sequence fitting (22 GCs from \cite{2017ApJ...838..162O}) and the RR Lyrae indicator (7 GCs from \cite{2016ApJ...827....2V, 2018ApJ...862...72V}, 16 GCs from \cite{2019ApJ...871...49H} plus M\,4 from \cite{2015ApJ...799..165B} and \cite{2015ApJ...808...11N}, M\,5 from \cite{2010A&A...516A..55C}, M\,22 from \cite{2013AJ....145...43K}, M\,62 from \cite{2010AJ....140.1766C}, NGC\,6362 from \cite{2018AN....339..183A} and NGC\,6723 from \cite{2014ApJS..210....6L}).

\begin{figure*} %[ht]
\vspace*{-3.0cm}\hspace*{-0.50cm} \includegraphics[width=2.3\columnwidth]{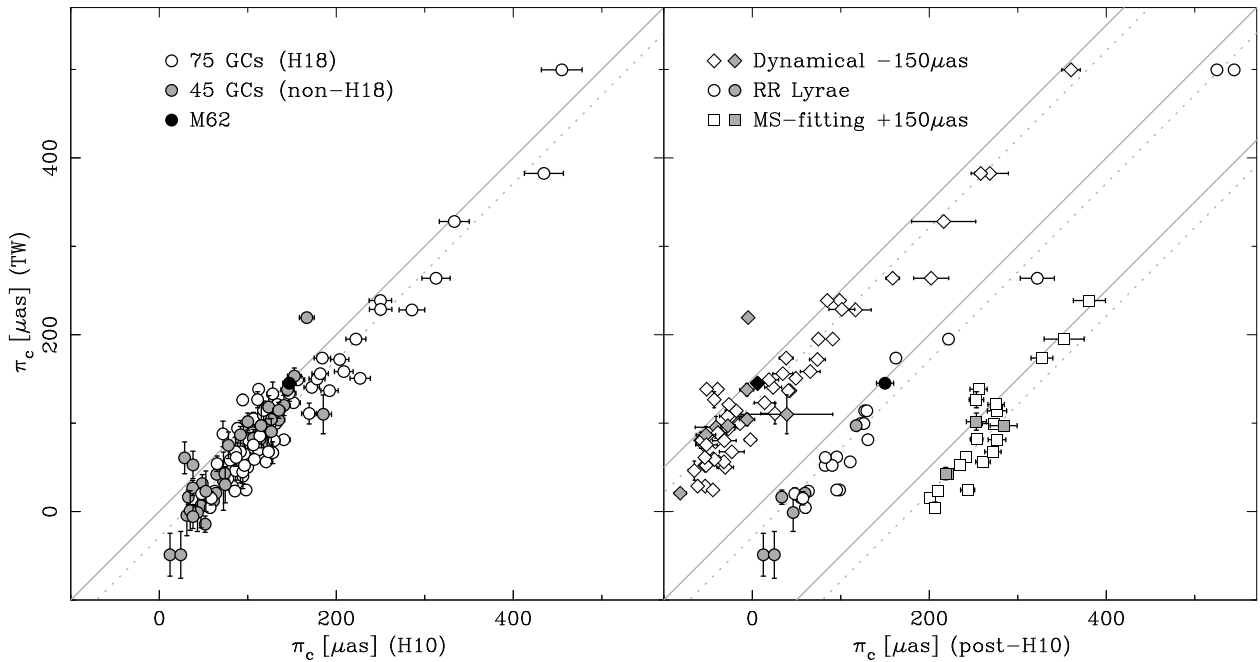}
\vspace*{-2.5cm}\caption{Comparisons of the parallaxes of GCs between this work (TW) and other indirect parallax results. Left panel: the comparison of H10. A typical error of 0.10 mag of the distance modulus is adopted for all H10 values. Hollow circles are the 75 common GCs of H18, and grey filled symbols are the additional 45 GCs resolved by the mixture model. Right panel: the comparison of selected indirect measurements after H10 (post-H10), with different methods plotted separately with different symbols and also shifted horizontally (see text for details of the sources). The black filled symbols are for M\,62 particularly. Solid and dotted lines represent the equal value and an offset of -29 $\mu$as. respectively} \label{Fig7}
\end{figure*}

As shown in figure \ref{Fig7}, the direct parallaxes of this work are obviously correlated to those indirect measurements for all sources, but with significant offsets and scatters that quantified by $\langle \Delta\varpi_{\rm c} \rangle$ and $\sigma_{\Delta\varpi_{\rm c}}$ respectively. These two values are the same as what we have introduced in the comparison of H18 in Sec.~\ref{sec:discussion1}. They have been calculated for comparisons of samples of H10, post-H10 and their combination separately, while for the data of H10, a 0.1 mag error of distance modulus are assumed for all GCs. The results are listed in table \ref{table2}. For a single cluster, if there are multiple sources or measurements, all collected $\varpi_{\rm c}$ are used in comparison.

Considering the uncertainties, all offsets and scatters of these three samples are consistent with each other. For the combined sample, totally 238 $\Delta\varpi_{\rm c}$ values between this work and the others, we have $ \langle \Delta\varpi_{\rm c} \rangle = 27.6\pm1.7$ $\mu$as and $ \sigma_{\Delta\varpi_{\rm c}}=22.8\pm1.3$ $\mu$as. We also find there is no evidence indicating that these two values are dependent on distance.

Since $\varpi_{\rm c,TW}$ is the apparent parallax, the offset and the scatter are reasonably supposed to be occurred by the \gaia parallax zero-point and its variation. Definitely, the offset is in good agreement with the global zero-point of $\sim -29$ $\mu$as, though it was reported that this value changes with different tracers \citep{2018A&A...616A..17A}. The scatter $\sigma_{\Delta\varpi_{\rm c}}$ is a little bit smaller than the value from the whole sky QSOs of $\sim 25$ $\mu$as. It seems that the \gaia calibration noise, which mainly causes the variation of zero-point, maybe not that much.

In sum, the results of this work confirm that $-29$ $\mu$as and $25$ $\mu$as are reasonable values of the zero-point and its variation of the current \gaia parallax. Then, if we use the global zero-point to correct the apparent $\varpi_{\rm c}$, a corresponding systematic error should also be considered.
%--
%--

\subsection{Distances of individual nearby clusters}\label{sec:discussion3}

Taking the systematic error of $\sigma_{\varpi_{\rm c,sys}}=25\,\mu$as into account, the parallaxes of distant GCs have large uncertainties currently. However, the total fractional uncertainties $f_{\rm tot} = (\sigma_{\varpi_{\rm c,TW}}^2 + \sigma_{\varpi_{\rm c,sys}}^2)^{1/2} / \varpi_{\rm c}$ of nearby clusters are still small, about a few or 10 percent. It is a comparable level of those of the indirect measurements, or even better than them. We then discuss several particular nearby GCs individually as below. In computing their distances, the global zero-point $-29$ $\mu$as are corrected for the parallaxes listed in table \ref{table1}, while the statistical and the systematic errors are derived separately.

\begin{description}
  \item[\bf M\,4 (NGC\,6121):] $d_{\rm TW}=1.892\pm0.004\pm0.090$ kpc. This value agrees well with $d_{\rm H18}=1.890\pm0.003 $ kpc that derived from their parallax with the same correction of zero-point.  Although it is the closest GC to the Sun, it has a large variation of previous distance measurements (from 1.7 to 2.2 kpc) with other indirect methods. This probably dues to its unusual foreground reddening. Now the direct parallax from \gaia greatly constrains the distance to the level of $\sim 5\%$ uncertainty, even though the systematic error is involved.
  \item[\bf NGC\,6397:] $d_{\rm TW} = 2.431\pm0.008\pm0.148 $ kpc and $d_{\rm H18}=2.456\pm0.004 $ kpc. The difference is small but still exceed 2$\sigma$ of their statistical errors. As the second closest GC to us, and also as one of the oldest and metal poor GCs, it has remarkable research importance. It also reaches in the scope of the HST parallax observation with $d_{\rm HST} = 2.39\pm0.07\pm0.10 $ kpc \citep{2018ApJ...856L...6B}. These direct parallaxes from different space facilities are consistent with each other, though they are derived through different approaches or with different sub-samples of cluster members. These values are slightly smaller than the previous values ($\sim 2.6$ kpc).  The total fractional uncertainty of our parallax is about $6\%$, which is better than most of the indirect measurements.
  \item[\bf $\omega$\,Cen (NGC\,5139):] It is the largest Milky Way GC and is rich in RR Lyrae variables. $d_{\rm TW}=6.031\pm0.055\pm0.909$ kpc and $d_{\rm H18}=6.549\pm0.047 $ kpc. Although these two values mismatched, both of them are larger than the previous result of $\sim 5.2$ kpc, which is only marginally consistent within the systematic errors of our result.
  \item[\bf 47\,Tuc (NGC\,104):] It is the second brightest globular cluster after $\omega$\,Cen and located in the direction close to the SMC. This cluster reaches the highest statistical precision of our resolved GCs with $f \sim 0.2\%$. It has $d_{\rm TW} =4.462\pm0.010\pm0.498$ kpc and $d_{\rm H18}=4.446\pm0.004 $ kpc. The distance is also in good agreement with \cite{2018ApJ...867..132C}'s result of $d=4.45\pm0.01\pm0.12 $ kpc with the pairwise method to avoid the zero-point problem. Here we notice that this super concordance is because they happened to find the \gaia parallax of SMC is $-25$ $\mu$as, which is very similar to the global zero-point we adopted.
  \item[\bf M\,62 (NGC\,6266):] It is one of the three most RR Lyrae rich GCs. Here we mention it because of the most serious divergence between the results of $d_{\rm TW}=5.744\pm0.102\pm0.825$ kpc and $d_{\rm H18}=4.037\pm0.059 $ kpc. Our value is much closer to the previous results of indirect method (from 6.4 to 6.8 kpc), but only in marginal agreement.
\end{description}

We also notice that for three of five GCs (NGC\,6397, $\omega$\,Cen and M\,62 ) discussed here, the divergences between results of the mixture model and the H18's membership approach are significant. It implies that there remains a large space in investigating the direct parallax methods. Either to modify the distribution of field stars in the mixture model or to improve the membership determination approach, are both worthwhile.

\section{summary}\label{sec:summary}

The mixture model is found to be practical in determining the mean parallax of GCs, though only parallax data of the \gaia DR2 are used. So it is the most independent method of direct parallax.

120 GCs, more than $80\%$ of the identified Milky Way GCs, are well resolved by the mixture model. They construct the largest direct parallax sample with high accuracy and acceptable precision. This work also demonstrates the value of the mixture model in determining the statistical properties of mixed distributions, such as the parallax in a cluster region with large observational errors of individual stars.

In comparing with previous indirect measurements, the offset $\langle\Delta\varpi_{\rm c}\rangle=-27.6\pm1.7$ $\mu$as confirms the global zero-point of \gaia parallax and the scatter $\sigma_{\Delta\varpi_{\rm c}}=22.8\pm1.3$ $\mu$as is also agree with the \gaia calibration noise, whereas it is the largest sample of GCs as the tracer to discuss this issue.

Currently, our uncertainties are dominated by systematic errors from the variation of the \gaia zero-point, and only a few nearest GCs such as M\,4 and NGC\,6397 can reach the comparable or even higher precision level than previous estimates. However, we believe that the situation should be greatly improved with further data release of the \gaia.

\section*{Acknowledgements}

The authors thank the anonymous referee for valuable suggestions. This work is supported by the National Natural Science Foundation of China (NSFC) under grants 11390373. This work has made use of data from the European Space Agency (ESA) mission {\it Gaia} (https://www.cosmos.esa.int/gaia), processed by the {\it Gaia} Data Processing and Analysis Consortium (DPAC,  https://www.cosmos.esa.int/web/gaia/dpac/consortium). Funding for the DPAC has been provided by national institutions, in particular the institutions participating in the {\it Gaia} Multilateral Agreement.

\label{lastpage}
\end{document}